\documentclass[aip,amsmath,amssymb,reprint]{revtex4-1}
\usepackage{graphicx}
\usepackage[utf8]{inputenc}
\usepackage[T1]{fontenc}
\usepackage{mathptmx}
\usepackage[english]{babel}

\usepackage{dcolumn}
\usepackage{bm}
\usepackage{mathtools} 
\usepackage{hyperref}
\usepackage{graphicx}
\usepackage{color}
\usepackage{xcolor}
\usepackage{svg}
\usepackage{subfig}

\usepackage[utf8]{inputenc}
\usepackage[T1]{fontenc}

\usepackage{booktabs}

\begin{document}

\title{Two-dimensional Cahn-Hilliard simulations for coarsening kinetics of spinodal decomposition in binary mixtures}

\author {Bj{\"o}rn K{\"o}nig}
 \affiliation{Helmholtz Institute Erlangen-N{\"u}rnberg for Renewable Energy,
  Forschungszentrum J{\"u}lich,\\
  F{\"u}rther Strasse 248, 90429 N{\"u}rnberg, Germany}%
 \affiliation{Department of Chemical and Biological Engineering, Friedrich-Alexander-Universit{\"a}t Erlangen-N{\"u}rnberg, F{\"u}rther Stra{\ss}e 248, 90429 N{\"u}rnberg, Germany}
 \author{Olivier J.J. Ronsin}
 \affiliation{Helmholtz Institute Erlangen-N{\"u}rnberg for Renewable Energy,
  Forschungszentrum J{\"u}lich,\\
  F{\"u}rther Strasse 248, 90429 N{\"u}rnberg, Germany}%
 \affiliation{Department of Chemical and Biological Engineering, Friedrich-Alexander-Universit{\"a}t Erlangen-N{\"u}rnberg, F{\"u}rther Stra{\ss}e 248, 90429 N{\"u}rnberg, Germany}
 \author{Jens Harting}
\email{j.harting@fz-juelich.de}
 \affiliation{Helmholtz Institute Erlangen-N{\"u}rnberg for Renewable Energy,
  Forschungszentrum J{\"u}lich,\\
  F{\"u}rther Strasse 248, 90429 N{\"u}rnberg, Germany}%
 \affiliation{Department of Chemical and Biological Engineering and Department of Physics, Friedrich-Alexander-Universit{\"a}t Erlangen-N{\"u}rnberg, F{\"u}rther Stra{\ss}e 248,\\ 90429 N{\"u}rnberg, Germany}
 
\date{\today}

\begin{abstract}
\label{sec:abstract}

The evolution of the microstructure due to spinodal decomposition in phase separated mixtures has a strong impact on the final material properties. In the late stage of coarsening, the system is characterized by the growth of a single characteristic length scale $L\sim C t^{\alpha}$. To understand the structure-property relationship, the knowledge of the coarsening exponent $\alpha$ and the coarsening rate constant $C$ is mandatory. Since the existing literature is not entirely consistent, we perform phase field simulations based on the Cahn-Hilliard equation. We restrict ourselves to binary mixtures using a symmetric Flory-Huggins free energy and a constant mobility term and show that the coarsening for off-critical mixtures is slower than the expected $t^{1/3}$-growth. Instead, we find $\alpha$ to be dependent on the mixture composition and thus from the morphology. Finally, we propose a model to describe the complete coarsening kinetics including the rate constant $C$.

\end{abstract}

\maketitle

\section{Introduction}
\label{sec:introduction}

When a binary mixture like a polymer solution is quenched in the thermodynamically unstable region of its phase diagram, the system undergoes phase separation via spinodal decompostion (SD).  The period until the phases separate and reach their equilibrium concentrations is called ``early stage'' of the demixing. The time needed for phase separation, denoted $t_{0}$ in the following, is material specific and can vary over decades. Then, the system further evolves towards the thermodynamic equilibrium by coarsening of the separated phases, whereby the energetic contribution due to the interfacial tension between the separated domains decreases progressively \cite{Otto_2012, bray, ratke_voorhees}. The characteristic length scale $L(t)$ of the system therefore increases over time. Directly after the phase separation, during the so-called ``intermediate stage'', the coarsening behavior is more sophisticated and there is no theory for it. The ``late stage'' of the evolution, at longer times, has been widely investigated and theoretically described. Ostwald\cite{ostwald} and later Lifshitz, Slyozov and Wagner\cite{LSW1, LSW2} formalized a theory (commonly known as LSW-theory) for coarsening of spherical precipitates for the limiting case of zero volume fraction. They predicted that the domains grow with time $t$ as $L\sim t^{1/3}$, the larger domains growing at the expense of the smaller ones. In general, such a power law can be written as
\begin{equation}
    \label{eq:ripening_allgemein}
         L(t)^{1/\alpha} - L(t_{0})^{1/\alpha} = C (t-t_{0}).
    \end{equation}
Here, $L(t_{0})=L_{0}$ is the characteristic length of the phase separated system at $t_{0}$, $\alpha$ will be called in what follows the ``coarsening exponent'' and $C$ the ``coarsening rate constant''.

The properties of a material blend are greatly affected by its morphology. Coarsening of the morphology is therefore expected to have a significant impact on the material properties. Hence, coarsening is a long lasting subject of interest, mostly investigated together with SD in binary systems with experimental methods\cite{barton_dynamics_1998_pmma, tork3,laxminayaran, peo2002}, numerical simulations\cite{sappelt_spinodal_1997, ahluwalia_phase_1999, HGC07, sheng_coarsening_2010, dai_computational_2016, MANZANAREZ_2017}, analytical models\cite{KohnOtto2002, novickcohen, Otto_2012} and still an active area of research \cite{voorhees_2017, voorhees_nature, banc_2019, andrews_simulation_2020, HigginsCabral2020, MITimperial}. Being able to predict the actual time-dependent average domain size or characteristic length scale $L(t)$ is crucial for understanding the morphology-property relationship. In particular, as indicated by Eq.~(\ref{eq:ripening_allgemein}), the exact knowledge of the coarsening exponent $\alpha$ is therefore of vast importance. 

A specific example is the formation of the photoactive layer in organic solar cells. Due to the polymeric nature of the donor materials, phase separation via SD and subsequent coarsening is likely to happen and is assumed to play an important role. During processing, both photo-active materials are dissolved in solution and deposited on a substrate. In the drying process and with the associated concentration increase in the solution, the mixture may become immiscible and undergo phase separation. Whether SD is beneficial\cite{kouijzer, wodo2018, longye_haraldade, SD_wichtig_AdeJoule} or undesired and a source of low efficiency of the solar cell\cite{brabec2short, gasparini2020BHJ} is still unclear. In any case, the size and morphology of the microstructure is crucial for both efficiency and stability of the photoactive layer\cite{brabec_latest}.

\begin{figure*}[ht]%
  \centering
  \subfloat[][]{\includegraphics[width=0.45\linewidth]{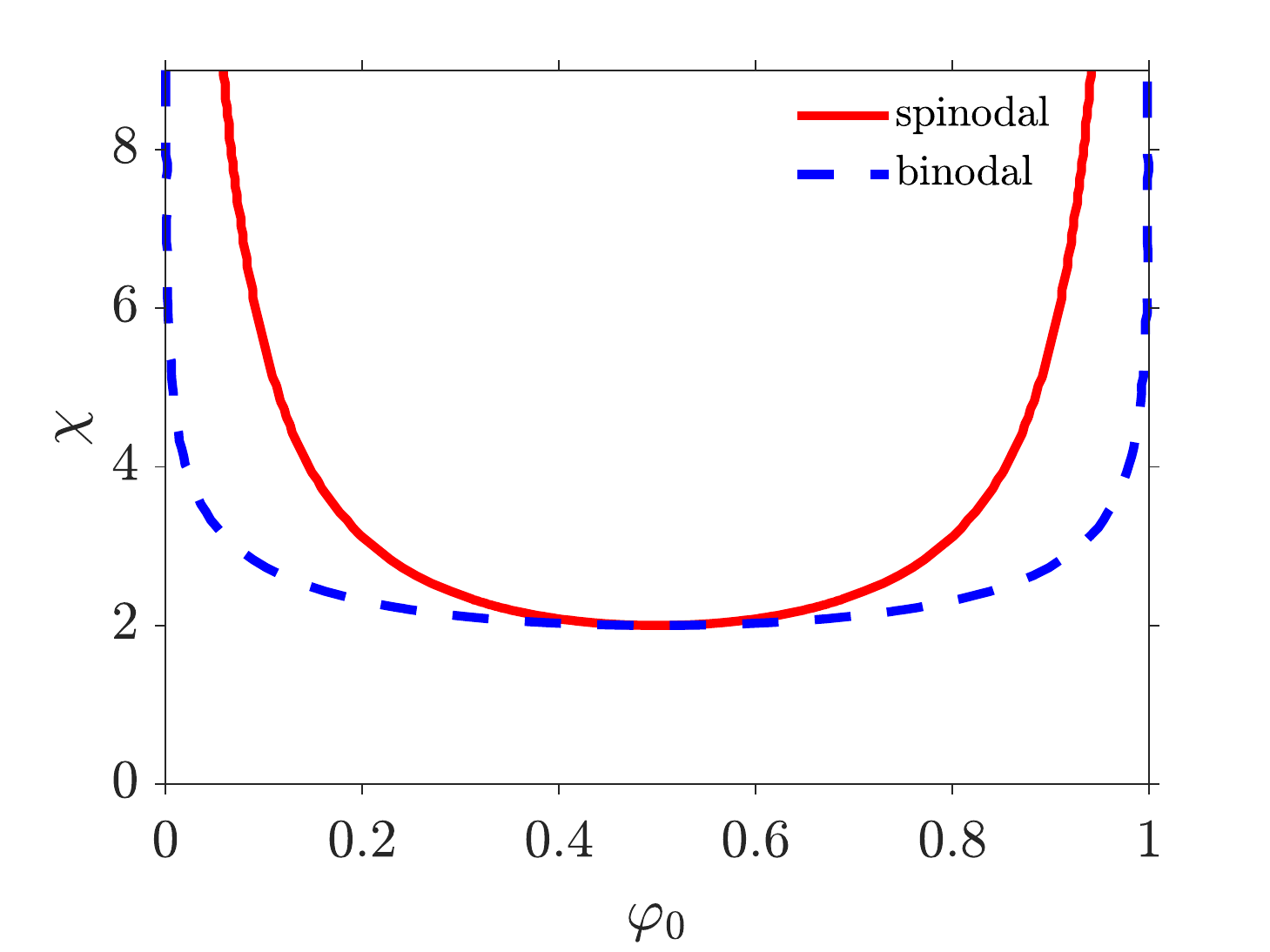}}%
  \qquad
  \subfloat[][]{\includegraphics[width=0.45\linewidth]{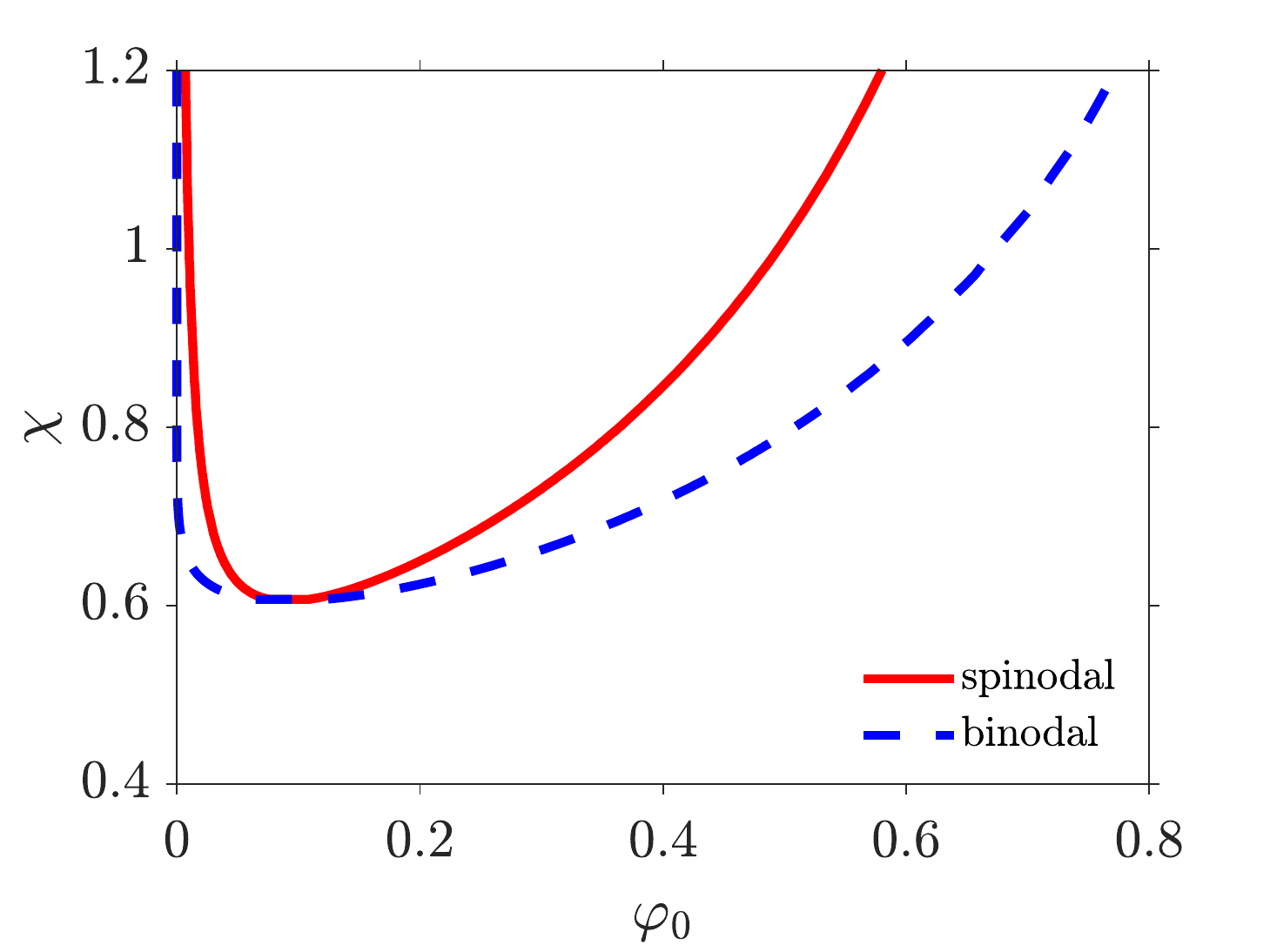}}%
  \caption{(a) Symmetric phase diagram for $N_{1}=N_{2}=1$ and (b) asymmetric phase diagram for $N_{1}=100$ and $N_{2}=1$. The spinodal line is calculated according to Eq.~(\ref{eq:spinodalline}).}%
  \label{fig:PD}
\end{figure*}

\renewcommand*{\arraystretch}{1.4}

\begin{table}[ht]
\centering
\begin{tabular}{c|c|c}
$\varphi_{0}$ & Quench depth & coarsening exponent $\alpha$\\
\hline

0.5 & deep & $1/3$ \cite{lacasta_domain_1992_critical,lacasta_domain_1992_offcritical, rogers_desai_1988critical, puri_phase-separation_1997, zhu_coarsening_1999, garcke_2003, sheng_coarsening_2010, dai_computational_2016, andrews_simulation_2020}  \\
0.45 & deep & $0.33$ \cite{rogers_desai_1989_offcritical}\\
0.4 & deep & $1/4$ \cite{garcke_2003}, $0.29$ \cite{rogers_desai_1989_offcritical}, $0.30$ \cite{chakra_brown_1993} \\
0.4 & shallow & $0.21$ \cite{chakra_brown_1993} \\
0.3 & deep & $1/3$ \cite{sheng_coarsening_2010}, $0.32$ \cite{lacasta_domain_1992_offcritical}, $0.29$ \cite{rogers_desai_1989_offcritical} \\
0.25 & deep & $1/3$ \cite{dai_computational_2016} \\
0.21 & deep & $1/3$ \cite{chakra_toral_gunton1995} \\
0.05 & deep & $1/3$ \cite{chakra_toral_gunton1995} \\
\end{tabular}
\caption{Dependencies of the coarsening exponent $\alpha$ using a symmetric phase diagram and a constant mobility $\Lambda$ - literature excerpt}
\label{tab:literature}
\end{table}


A widely used model for phase separation due to SD and subsequent coarsening is the Cahn-Hilliard equation~\cite{CHequation1, CHequation2}. It is a classical diffusion equation, but where the driving force for the system evolution is the gradient of the exchange chemical potential, which includes not only mixing terms but also surface tension terms in order to handle separated phases. 

In a binary blend like the ones investigated in this article, the system is described by the time- and space-dependent volume fraction field $\varphi$ of one of the two materials, while the other one is given by $1-\varphi$. The initial volume fraction is denoted by $\varphi_{0}$. For polymeric solutions, the Gibbs free energy of mixing is often expressed using the Flory-Huggins equation\cite{FH_equation}. It allows to take into account different molar volumes, often denoted as degree of polymerization $N_{i}$ of the $i^{th}$-mixture component, as well as different interaction parameters between each pair of materials. Equal degrees of polymerization result in a phase diagram which is symmetric about $\varphi_{0}=0.5$ as depicted in Fig.~\ref{fig:PD}(a), denoted as a symmetric blend or symmetric material system. Differences in $N_{i}$ lead to a strongly asymmetric phase diagram as illustrated in Fig.~\ref{fig:PD}(b). Besides the molar volumes and the initial volume fraction or initial composition of the mixture $\varphi_{0}$, a binary blend is characterized by the interaction parameter $\chi$ which describes the degree of incompatibility of the materials. Focusing on the unstable region of the phase diagram, the degree of incompatibility will be called quench depth in this paper. Almost pure phases are characteristic for deep quenches. The mixture composition with the lowest value of $\chi$ which will lead to SD is defined as the critical composition, while other mixture compositions are denoted as off-critical. 
For a symmetric blend, Fig.~\ref{fig:PD}(a), a mixture with equal volume fractions (denoted in this article as a 50/50-mixture or a mixture with initial composition of $\varphi_{0}=0.5$) constitutes the critical mixture.

An essential property in the Cahn-Hilliard model is the mobility, which describes how fast the system reacts to the thermodynamic driving force and is characterized by the mobility coefficient $\Lambda$. Beyond the simplest assumption of a composition-independent constant value for $\Lambda$, there exist many other composition-dependent mobility models in the literature\cite{sheng_coarsening_2010,dai_computational_2016,andrews_simulation_2020}.

A lot of research has already been done on simulations of the Cahn-Hilliard equation in 2D using different setups. Table~\ref{tab:literature} is a selection that summarizes to the best of our knowledge the key references for coarsening kinetics using 2D numerical simulations of the Cahn-Hilliard equation under the assumption of a symmetric phase diagram and a constant mobility $\Lambda$. It can be seen that the coarsening exponent $\alpha$ for the critical composition is found very robust to be $1/3$, while the results for $\alpha$ for off-critical mixtures are inconsistent. 
Additionally, the coarsening kinetics under the assumption of composition-dependent mobility functions have been reported. Using a parabolic or double-degenerate mobility, Refs.\cite{sheng_coarsening_2010} and \cite{dai_computational_2016} report a $\varphi_{0}$-independent $\alpha$ equal to $1/4$. For a one-sided mobility, the authors of Ref.\cite{sheng_coarsening_2010} obtain $\alpha=1/3.3$, while in Ref.\cite{dai_computational_2016} $\alpha$ is reported to be $\varphi_{0}$-dependent and found $1/3$ (25/75 and 50/50) and $1/4$ (75/25). For this mobility and $\varphi_{0}=0.5$, Andrews et al.\cite{andrews_simulation_2020} report a time-dependent coarsening rate constant $C=C(t)$.

While symmetric or nearly symmetric phase diagrams are often encountered in metallic alloys, polymeric solutions are typically characterized by highly asymmetric phase diagrams. For these systems, some recent articles\cite{Mino2015, ASYNC_erreichen, Nestler2019, Nestler2020} reveal novel details for the microstructure formation in the early stages of SD. However, the literature on coarsening kinetics is scarce in comparison to the symmetric case\cite{MANZANAREZ_2017, barton_dynamics_1998_SDnearTG}.

Despite the numerous already available theoretical studies listed above, we believe that a systematic and complete overview of the coarsening kinetics is still missing. A systematic investigation should basically differentiate between the effect of quenching conditions ($\varphi_{0}$ and $\chi$) and the effect of the assumption for the mobility term $\Lambda$, both for symmetric as well as asymmetric material systems.

In this work, we perform large scale simulations of the Cahn-Hilliard equation in 2D and systematically investigate the kinetics of SD. Section~\ref{sec:equations} describes the used phase field framework in detail. In Sec.~\ref{sec:scalinglaws}, we investigate the early stages of SD and check the validity of the well-known scaling laws for the time until demixing $t_{0}$, and the corresponding initial average domain size of the phase separated structure $L_{0}$, over a broad range of parameters. In Sec.~\ref{sec:coarsening}, we investigate the coarsening behavior during the late stages of the phase separation, focusing on a symmetric material system ($N_{1}=N_{2}=1$) and a constant, composition-independent mobility coefficient $\Lambda$. There, we investigate how the coarsening kinetics may vary with the quench depth and the blend composition. Finally, we give a short summary of our findings and conclude in Sec.~\ref{sec:conclusion}.

\section{Simulation model and implementation}
\label{sec:equations}

The kinetic evolution of the system towards its thermodynamic equilibrium is simulated with the phase-field framework proposed in our previous papers\cite{olivier1, olivier2}. 

As a starting point, we describe the system with the help of its free energy functional which defines its thermodynamic properties. Without loss of generality, the free energy functional reads

      \begin{equation}
      \label{eq:freeenergyfunctional}
        G^{\mathrm{tot}}  = \int_{V} \ (G^{\mathrm{loc}}+G^{\mathrm{nonloc}}) \ \mathrm{d}V  \, ,
    \end{equation} 
where $V$ is the system volume, $G^\mathrm{loc}$ is the local free energy density and $G^\mathrm{nonloc}$ the non-local contribution due to the field gradients. The local part of the free energy corresponds to the Flory-Huggins theory of mixing\cite{FH_equation} and an additional purely numerical contribution
 
    \begin{align}
        \label{eq:localfreeenergy_detailed}
        G^\mathrm{loc}  = \frac{RT}{v_{0}}
        \left(
        \ \frac{\varphi\ln{(\varphi)}}{N_{1}}+\frac{(1-\varphi)\ln{(1-\varphi)}}{N_{2}}
        \right)\nonumber\\
        +\varphi \ (1-\varphi) \chi
        +{k} \left(\frac{1}{\varphi}+\frac{1}{1-\varphi}\right).
    \end{align} 
In the equation above, $\varphi$ is the volume fraction of the first material, $R$ is the gas constant, $T$ the temperature, and $v_0$ the molar volume of the lattice site in the Flory-Huggins theory. The molar volume of the material $i$ is given by $v_i=N_iv_0$, while $\chi$ is the Flory-Huggins interaction parameter. The first term in Eq.~(\ref{eq:localfreeenergy_detailed}) corresponds to the entropy of mixing, the second one to the enthalpic interactions and the last one is a purely numeric correction which supports numerical stability, especially for high values of $N\chi$ for which the binodal concentrations can be very close to 0 an 1.

The non-local part of the free energy functional describes the contributions of the interfacial tension with
     \begin{equation}
     \label{eq:nonlocalfreeenergy}
        G^\mathrm{nonloc}  = \frac{1}{2} 
        \ \kappa \ (\nabla \varphi)^{2},
    \end{equation}  
where $\kappa$ is the surface tension parameter. 

The phase diagram can be computed from the local free energy. As an example, the phase diagram for a symmetric material system with $N_{1}=N_{2}=1$ is given in Fig.~\ref{fig:PD}(a), while a phase diagram where one material has a much higher molar volume than the other one ($N_{1}=100, N_{2}=1$) is shown in Fig.~\ref{fig:PD}(b). The spinodal line (red full line) defines the instability region of the blend, above which it is unstable and undergoes spontaneous SD. The equation of the spinodal line $\chi_{s}$ is given by \cite{rubinstein} 
    \begin{equation}
    \label{eq:spinodalline}
         \chi_{s}=\frac{1}{2} \left( \frac{1}{N_1 \varphi}+ \frac{1}{N_2 (1-\varphi)} \right).
    \end{equation} 

The composition of the phases after demixing are given by the binodal line (blue dashed line). In the region between the binodal and the spinodal line, the system is metastable and can demix by nucleation and growth of isolated droplets. The binodal concentrations can be computed by the equality of the chemical potential in both phases for both materials (common tangent construction\cite{tangentconstruction}). This can be solved analytically in the symmetric case but only numerically in the asymmetric case.

The kinetic equation for the volume fraction $\varphi$, which is a conserved quantity, is based on the formalism initially proposed by Cahn and Hilliard\cite{CHequation1, CHequation2}:
    \begin{equation}
    \label{eq:to_solve_equation}
         \frac{\partial \varphi}{\partial t}=\frac{v_{0}}{RT}\nabla \left(
         \Lambda \nabla \mu^{ex} \right)
    \end{equation}
This can be understood as a continuity equation, the fluid flux being proportional to the gradient of the exchange chemical potential density $\mu^{ex}$, which is the driving force for the system evolution. $\Lambda$ is the mobility coefficient related to the diffusional properties of both materials. The exchange chemical potential $\mu^{ex}$ is calculated from the free energy functional in the following way:   
    \begin{align}
    \label{eq:exchangechemicalpotential}
         \mu^{ex}=
         \frac{\partial G}{\partial \varphi}-
         \nabla \frac{\partial G}{\partial (\nabla \varphi)} 
    \end{align}
The first term corresponds to the chemical potential, whereas the second contribution takes into account the potential due to concentration gradients. The Cahn-Hilliard equation ensures that the system progressively minimizes its free energy relative to the volume fraction, i.e.~it relaxes towards its thermodynamic equilibrium.
 
The importance of the mobility $\Lambda$ for the kinetic evolution of the blend, and especially its dependence on the composition variable $\varphi$ has already been intensively investigated by many authors, as stated in Sec.~\ref{sec:introduction}. In Cahn’s original assumption, the flux is simply proportional to the gradient of the generalized exchange chemical potential through a constant mobility coefficient. However, the mobility has to depend on the local mixture composition in order to ensure the incompressibility constraint together with the Gibbs-Duhem relationship. Several theories have been proposed to derive correct expressions for the coupled fluxes in multinary mixtures, among which the ``slow mode theory''\cite{SM} and the “fast-mode theory” \cite{FM} are the most successful. Their names come from the fact that the mutual diffusion coefficient derived from the Cahn-Hilliard equation in a binary system is controlled by the slowest component in the ``slow-mode theory'', while it is controlled by the fastest component in the ``fast-mode theory''. For a binary system, the mobilities for the fast mode theory $\Lambda_{FM}$ and for the slow mode theory $\Lambda_{SM}$ respectively read 
    \begin{equation}
    \label{eq:FM_equation}
         \Lambda_{FM} = \varphi (1-\varphi)^{2}N_{1}D_{1} + \varphi^{2} (1-\varphi)N_{2}D_{2},
    \end{equation}
    \begin{equation}
    \label{eq:SM_equation}
         \Lambda_{SM} = \frac{\varphi(1-\varphi) N_{1}D_{1} N_{2}D_{2}}{\varphi N_{1}D_{1} +(1-\varphi)N_{2}D_{2}}.
    \end{equation}
 In Eqns.~(\ref{eq:FM_equation}) and (\ref{eq:SM_equation}), $D_{1}$ and $D_{2}$ are the self-diffusion coefficients of both materials, which in general also depend themselves on the composition of the mixture. Various phenomenological models exist in the literature in order to describe the composition dependence of the diffusion coefficients, among others the equation initially proposed by Vignes\cite{LBV_Vignes} and the one proposed by Liu, Bardow and Vlugt (LBV)\cite{LBV_Vignes}. 
 For both of these models, the composition-dependent diffusion coefficients can be calculated with the help of the self-diffusion coefficients in the pure materials, with $D_{ij}$ being the self-diffusion coefficient of material $i$ in a matrix of $100\%$ material $j$. Hence, for a binary mixture, only four self-diffusion coefficients ($D_{11},D_{12},D_{21},D_{22}$) need to be supplied and $D_{1}$ and $D_{2}$ are given as $D_{1}(\varphi)=f(D_{11},D_{12},\varphi)$ and $D_{2}(\varphi)=f(D_{22},D_{21},\varphi)$. For the binary case, the LBV-assumption reads
     \begin{equation}
    \label{eq:Dconnection1}
         D_{1} = \frac{1}{\frac{\varphi}{D_{11}}+\frac{(1-\varphi)}{D_{12}}},
    \end{equation}
 while the Vignes-model is given by 
      \begin{equation}
    \label{eq:Dconnection2}
         D_{1} = D_{11}^{\varphi} D_{12}^{(1-\varphi)}.
    \end{equation}
The same holds true for $D_{2}$ with $D_{21}$ and $D_{22}$. Furthermore, a linear weight of $D_{11}$ and $D_{12}$ can also be used,   

     \begin{equation}
    \label{eq:Dconnection3}
         D_{1} = D_{11}\varphi+D_{12}(1-\varphi).
    \end{equation}
This renders the final expression of the mobility $\Lambda$ complicated in the general case. However, if we consider the case of a symmetric material system ($N_{1}=N_{2}=1$) and assume that the system can be described by one single, constant composition-independent self-diffusion coefficient $D$, both the fast-mode and the slow-mode model reduce to the same expression $\Lambda=D\varphi(1-\varphi)$ which is often referred to as the "double degenerate" mobility in the literature. This mobility is used as a model for surface diffusion driven coarsening\cite{sheng_coarsening_2010, dai_computational_2016}. It emphasizes the diffusion in the interface regions between the bulk phases. Nevertheless, one needs to keep in mind that these simplifications are not suitable to simulate polymeric material systems which are characterized by highly dynamic asymmetries.

The Cahn-Hilliard equation (\ref{eq:to_solve_equation}) is written in the split form \cite{Elliott}, discretized with finite volumes, and solved using an implicit backward Euler finite difference scheme. To this end, the local part of the free energy is linearized consistently. The code is parallelized and the linear system is solved using PETSc\cite{petsc-web-page, petsc-user-ref, petsc-efficient} with an iterative solver (GMRES method together with a bloc Jacobi preconditioner \cite{gmres}). We make use of the unconditional stability of the Euler backward scheme to achieve large time steps. Hereby, we use an adaptive time stepping strategy similar to the one described by Wodo and Ganapathysubramanian\cite{wodo2012}, based on the number of iterations required for the convergence of the solver. We perform 2D simulations with periodic boundary conditions in both directions. The phase separation is initiated with the help of a small initial perturbation to the homogeneous volume fraction field. The grid resolution is adapted to each parameter set in order to discretize the interface between separated phases with at least five grid points. With the chosen parameters, the interface thickness lies typically between $5~nm$ and $50~nm$ so that the grid spacing lies between $1~nm$ and $8~nm$. It has been verified that all simulations are numerically converged in both time and space.

\section{Early stage behavior}
\label{sec:scalinglaws}

\renewcommand*{\arraystretch}{1.5}

\begin{table}[ht]
    \centering
    \begin{tabular}{cc}
         \hline
         T & $300$ K  \\
         $\nu_{0}$ & $10^{-3} m^{3} mol^{-1}$ \\
         $\rho_{i}$ & 1000 $kg m^{-3}$ \\
         $k$ & $10^{-5} J m^{-3}$ \\
         $N_{1}$ & 1-100 \\
         $N_{2}$ & 1 \\
         $\chi$ & 0.6-10 \\
         $\kappa_{i}$ & $10^{-8}-10^{-12} J m^{-1}$ \\
         $D_{ij}$ & $10^{-9}-10^{-16} m^{2} s^{-1}$ \\
         $\Lambda$ & constant, Eqns.~(\ref{eq:FM_equation}) and (\ref{eq:SM_equation}) \\
         \hline
    \end{tabular}
    \caption{Simulation parameters for investigation of the early stage behavior in Sec.~\ref{sec:scalinglaws}}
    \label{tab:simulationparameters1}
\end{table}

    \begin{figure}[ht]
        \centering
        \includegraphics[scale = 0.58]{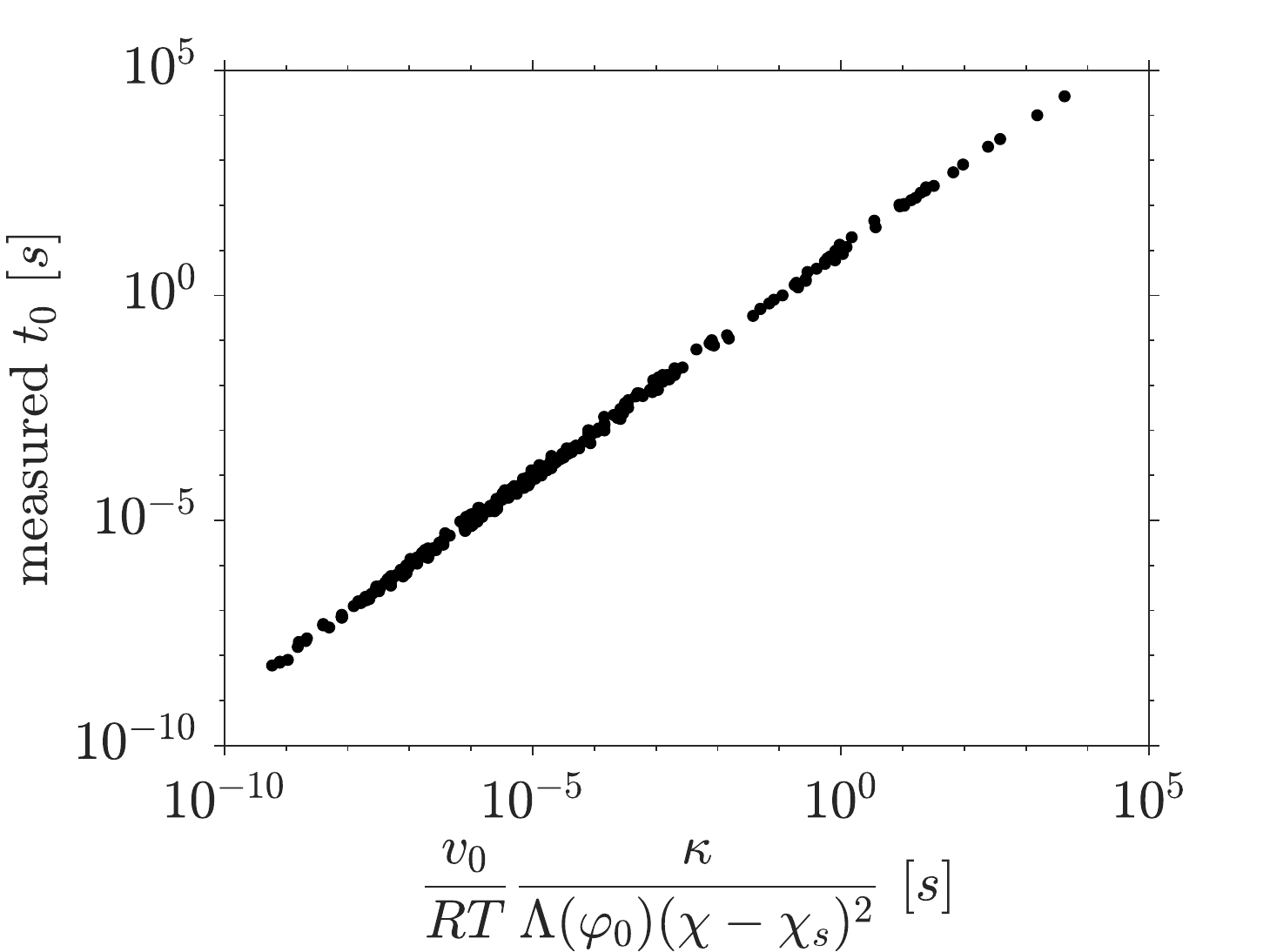}
        \caption{Simulated time until phase separation $t_{0}$ versus expected time from Eq.~(\ref{eq:lsa7}). We determine the prefactor $a$ to be 9.7.}
        \label{fig:t0scalinglaw}
    \end{figure}
    
    \begin{figure}[ht]
        \centering
        \includegraphics[scale = 0.58]{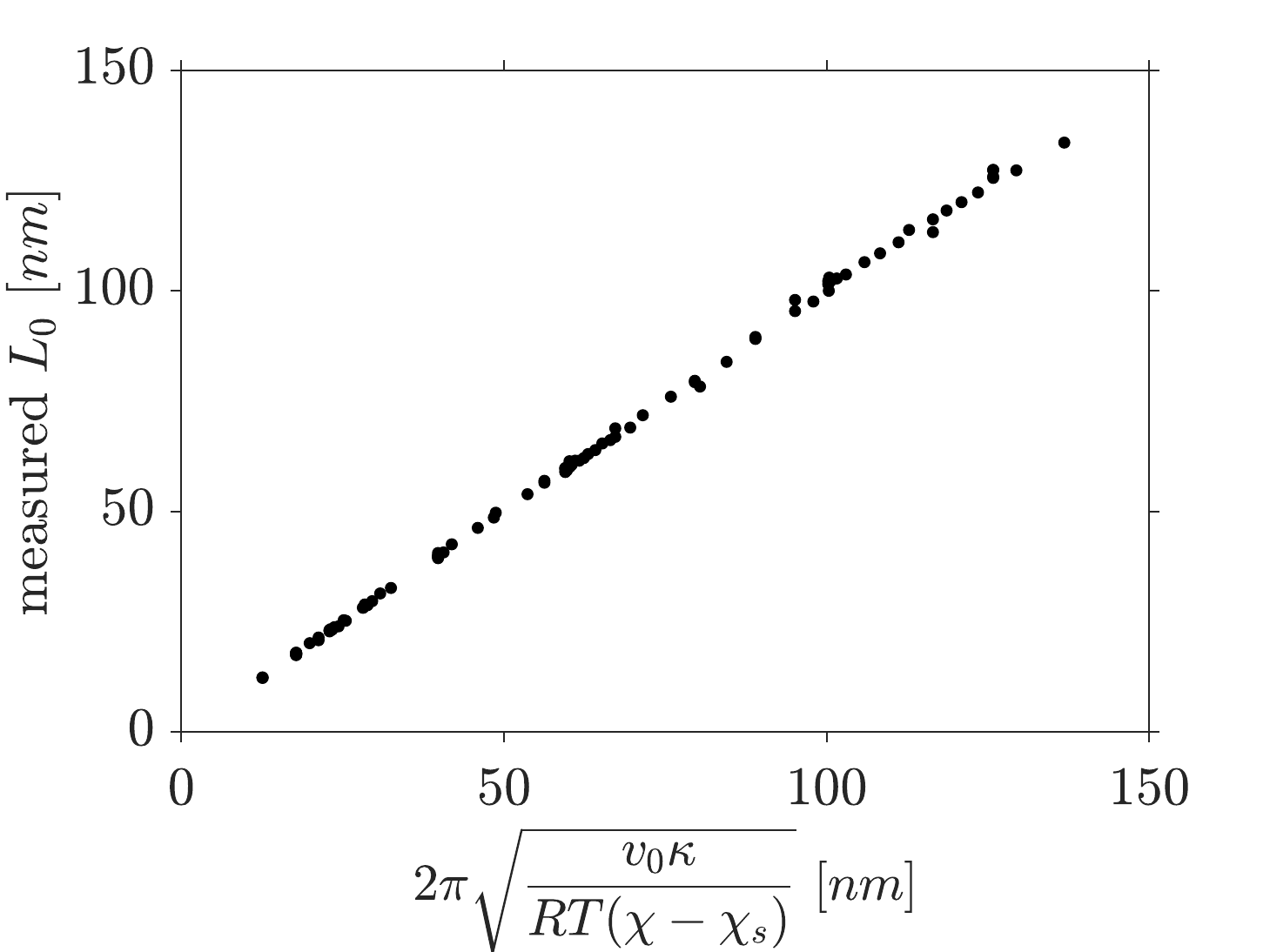}
        \caption{
         Simulated initial characteristic size $L_{0}$ versus expected size from Eq.~(\ref{eq:lsa6}).}
        \label{fig:R0scalinglaw}
    \end{figure}

From the linear analysis of the Cahn-Hilliard equation  the time needed for the SD to take place $t_{0}$, and the initial characteristic size $L_{0}$ of the separated phases can be calculated as\cite{HigginsCabral2018, HigginsCabral2020}
    \begin{equation}
        t_{0} = a \frac{v_{0}}{RT} \ \frac{\kappa}{\Lambda(\varphi_{0}) \ (\chi-\chi_{s})^{2}}
    \label{eq:lsa7}
    \end{equation} 
and    
    \begin{equation}
        L_{0} = 2 \pi \sqrt{\frac{v_{0}\kappa}{RT(\chi-\chi_{s})}}.
    \label{eq:lsa6}
    \end{equation} 
Here, $a$ is an arbitrary constant which depends on the definition taken for $t_{0}$. In this work, $t_{0}$ is determined as the time required for the first phase to reach the equilibrium binodal composition within an error of $1\%$. As a measure for the average domain size or the characteristic length scale $L(t)$, we calculate the 2D-structure factor from the Fourier transform of the volume fraction field. We obtain the probability distribution $p(q,t)$ of $q$-vectors by integration over all directions. Taking the inverse of the mean value of $q$ over this distribution gives the characteristic length scale as
    \begin{equation}
        L(t)= 2\pi/
        \left(\int qp(q,t)dq
        \right).
    \label{eq:characteristiclength}
    \end{equation} 

Both theoretical predictions for $t_{0}$ and $L_{0}$ given by Eqns.~(\ref{eq:lsa7}) and (\ref{eq:lsa6}) are compared to numerous simulations were we vary the parameters over a broad range according to Tab.~\ref{tab:simulationparameters1}. Thereby, we also use sophisticated mobility functions given by Equation (\ref{eq:FM_equation}) and \ref{eq:SM_equation} in combination with composition-dependent self-diffusion coefficients (Eqns.~(\ref{eq:Dconnection1}) to (\ref{eq:Dconnection3})) which themselves differ by decades. 
The results, shown in Figs.~\ref{fig:t0scalinglaw} and \ref{fig:R0scalinglaw}, show an excellent agreement with the predictions and validate our numerical implementation. These results are also useful for the evaluation of the late stage coarsening, since $t_{0}$ and $L_{0}$ are required in order to determine $L(t)$ according to Eq.~(\ref{eq:ripening_allgemein}). 
In addition, the form of our equations for $t_{0}$ and $L_{0}$ allows to directly identify the impact of parameter variations. As illustrated in Figs.~\ref{fig:t0scalinglaw} and \ref{fig:R0scalinglaw}, $t_{0}$ and $L_{0}$ may change by orders of magnitude. For a quantitative simulation of realistic material systems, this emphasizes the importance of the correct choice of the simulation parameters as well as the mobility assumption.

\section{Late Stage Coarsening kinetics}
\label{sec:coarsening}

    \renewcommand*{\arraystretch}{1.5}
    
    \begin{table}[ht]
    \centering
    \begin{tabular}{cc}
         \hline
         T & $300$ K  \\
         $\nu_{0}$ & $10^{-3} m^{3} mol^{-1}$ \\
         $\rho_{i}$ & 1000 $kg m^{-3}$ \\
         $k$ & $10^{-5} J m^{-3}$ \\
         $N_{1}=N_{2}$ & 1 \\
         $\chi$ & see Fig.~\ref{fig:testedconfigurations} \\
         $\kappa_{i}$ & $2\cdot10^{-10}J m^{-1}$ \\
         $D$ & $10^{-10}m^{2} s^{-1}$ \\
         $\Lambda$ & $D$ (constant) \\
         \hline
    \end{tabular}
    \caption{Simulation parameters for investigation of the late stage coarsening kinetics in Sec.~\ref{sec:coarsening}}
    \label{tab:simulationparameters2}
    \end{table}

    \begin{figure}[ht]
        \centering
        \includegraphics[scale = 0.58]{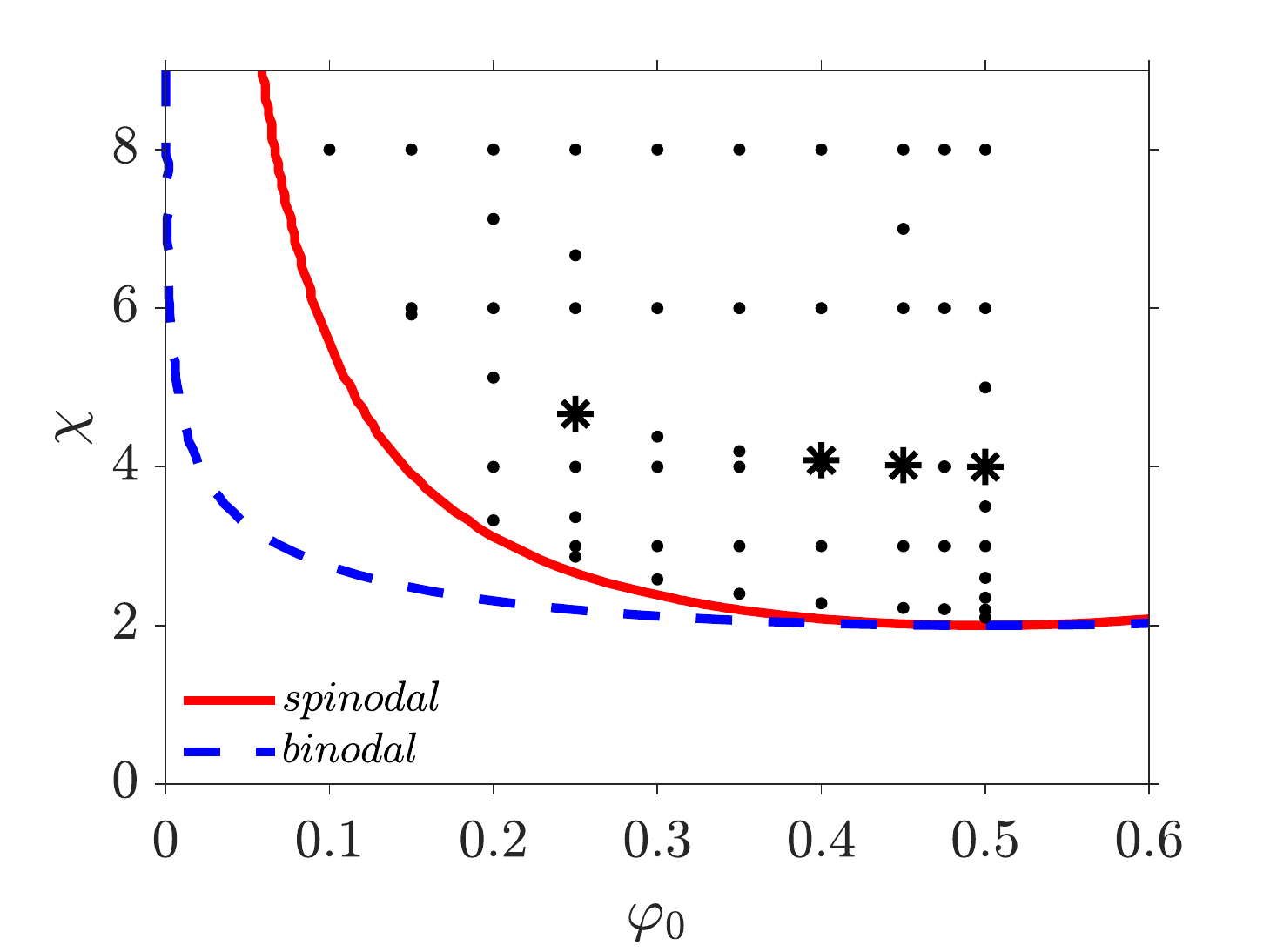}
        \caption{Tested quenching conditions in the spinodal region for the late stage coarsening kinetics. The results for the asterisk-marked conditions are presented in Figs.~\ref{fig:morpho} and \ref{fig:kstudy_lengthscale}.} \label{fig:testedconfigurations}
    \end{figure}


    \begin{figure*}[ht]
        \centering
         \includegraphics[scale = 0.78]{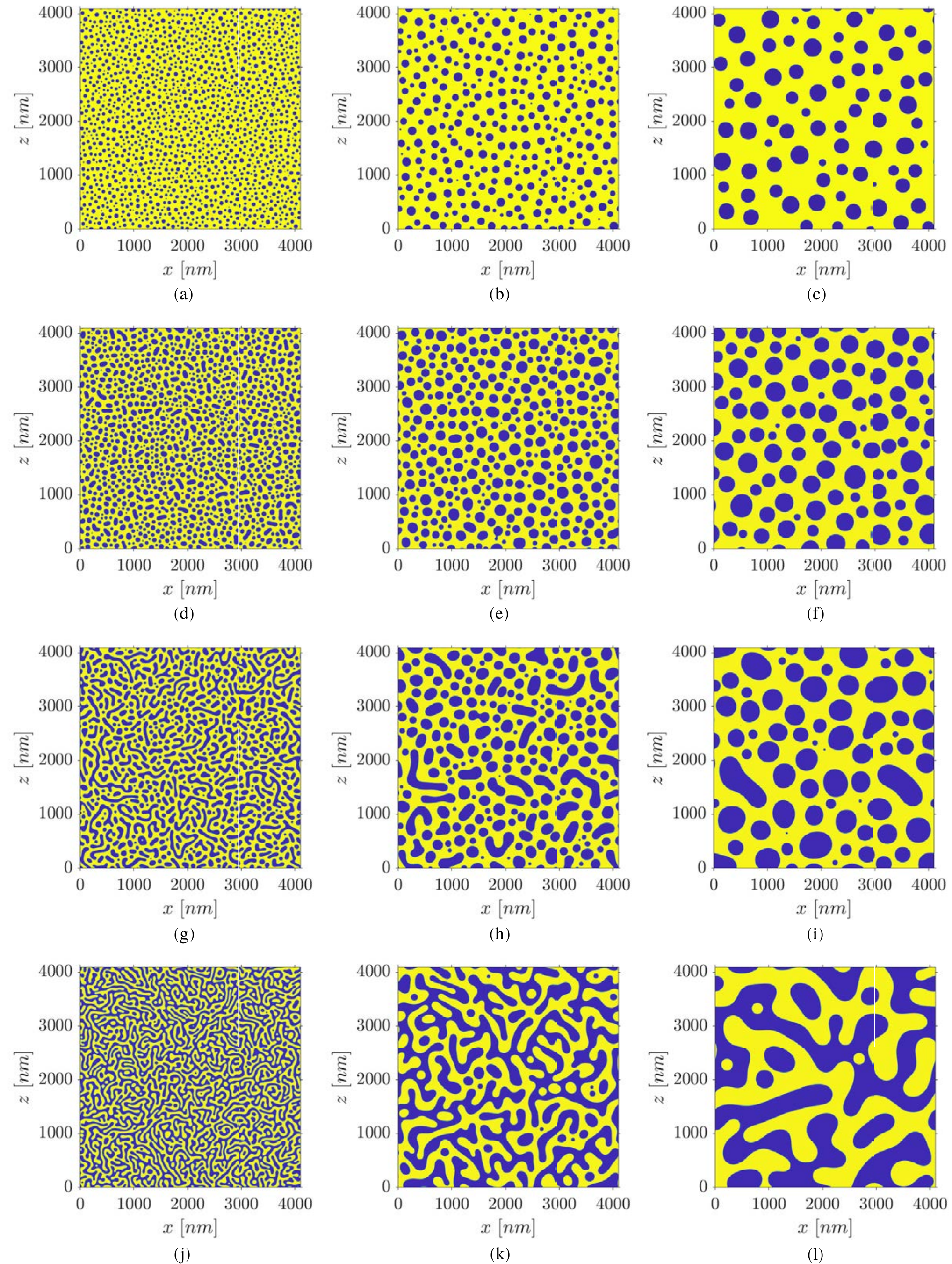}
        \caption{Resulting morphologies for the asterisk-marked quench points in Fig.~\ref{fig:testedconfigurations} for $\varphi_{0}=0.25$ ((a)-(c)), $\varphi_{0}=0.40$ ((d)-(f)), $\varphi_{0}=0.45$ ((g)-(i)) and $\varphi_{0}=0.5$ ((j)-(l)). The 1st column corresponds to $4\times10^{-5} \ s$, the 2nd column to $4\times10^{-4} \ s$ and the 3rd one to $4\times10^{-3} \ s$. For $\varphi_{0}=0.25$ to $0.45$, the majority concentration of the mixture constitutes the yellow matrix phase.}
        \label{fig:morpho}
    \end{figure*}

    \begin{figure}[ht]
        \centering
         \includegraphics[scale = 0.58]{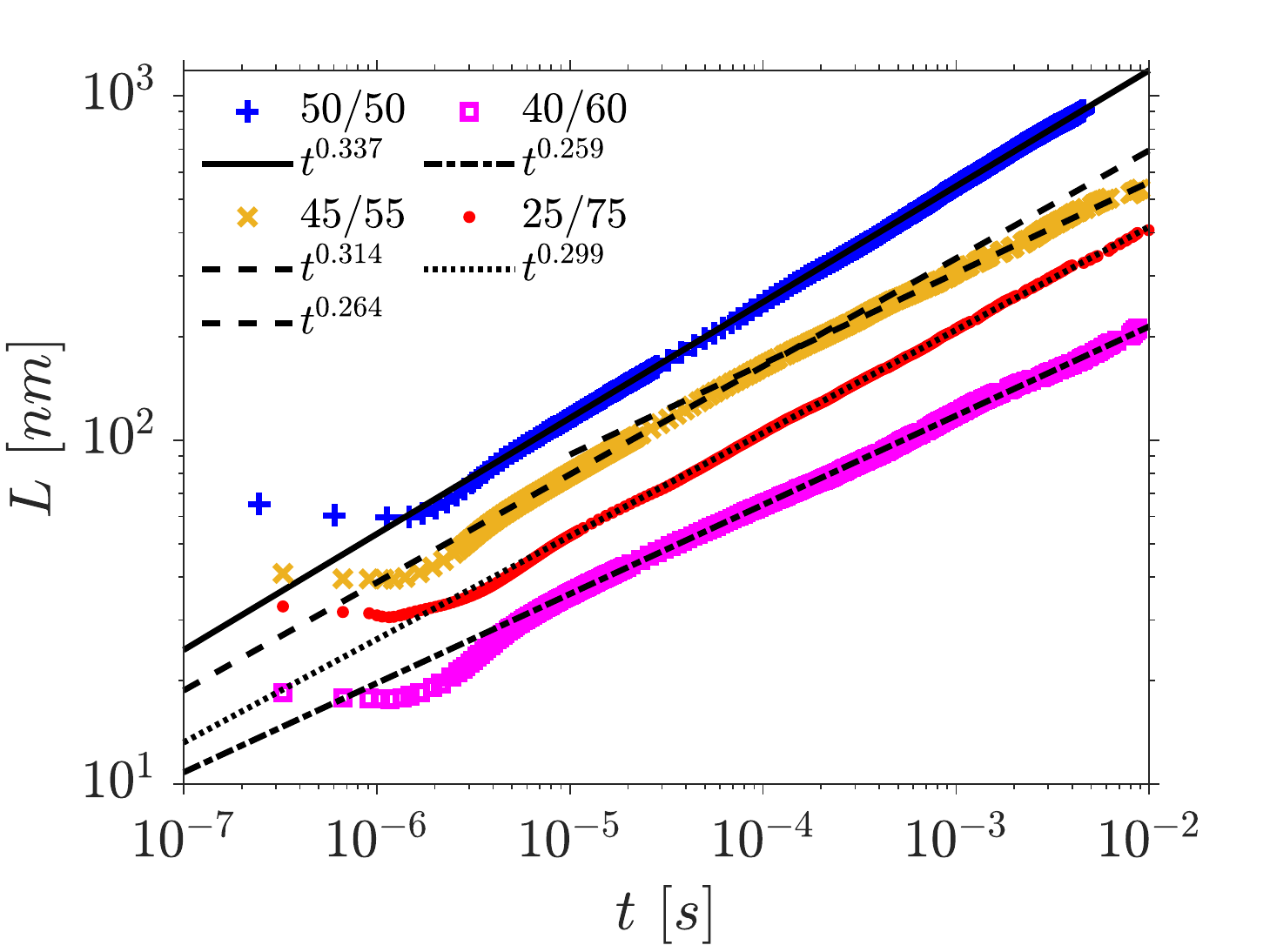}
        \caption{
        Characteristic length scales $L$ as a function of time $t$ for $\varphi_{0}=0.5$ (blue plus signs), $\varphi_{0}=0.45$ (ochre crosses), $\varphi_{0}=0.4$ (magenta squares) and $\varphi_{0}=0.25$ (red points) corresponding to Fig.~\ref{fig:morpho} and power law reference lines $t^{\alpha}$. Note that we multiplied $L(t)$ for $\varphi_{0}=0.50$, $0.40$ and $0.25$ by arbitrary constants for a better visualization.}
        \label{fig:kstudy_lengthscale}
    \end{figure}

In this Section, we investigate the late stage coarsening kinetics of a binary symmetric material system for various quenching conditions. The mobility $\Lambda$ is assumed to be constant. We characterize the coarsening kinetics by fitting the characteristic length scale $L(t)$ obtained from the simulations according to Eq.~(\ref{eq:characteristiclength}), with the growth law, Eq.~(\ref{eq:ripening_allgemein}). It is our aim to identify the coarsening exponent $\alpha$ and the coarsening rate constant $C$. We perform 2D simulations with a grid size of $1024 \times 1024$ elements. The large grid size ensures that the late stage coarsening lasts for at least two decades of physical time with a significant number of phase-separated domains. This is necessary for a precise evaluation of $\alpha$ and $C$. The tested compositions and quench depths are presented in Fig.~\ref{fig:testedconfigurations}.  We vary the composition $\varphi$ at fixed interaction parameter $\chi$, the interaction parameter $\chi$ at fixed composition $\varphi$, and both together at fixed quench depth $\chi-\chi_{s}$. Altogether, we investigate 60 different conditions covering a wide range of the unstable region.

    \begin{figure}[ht]
        \centering
        \includegraphics[scale = 0.58]{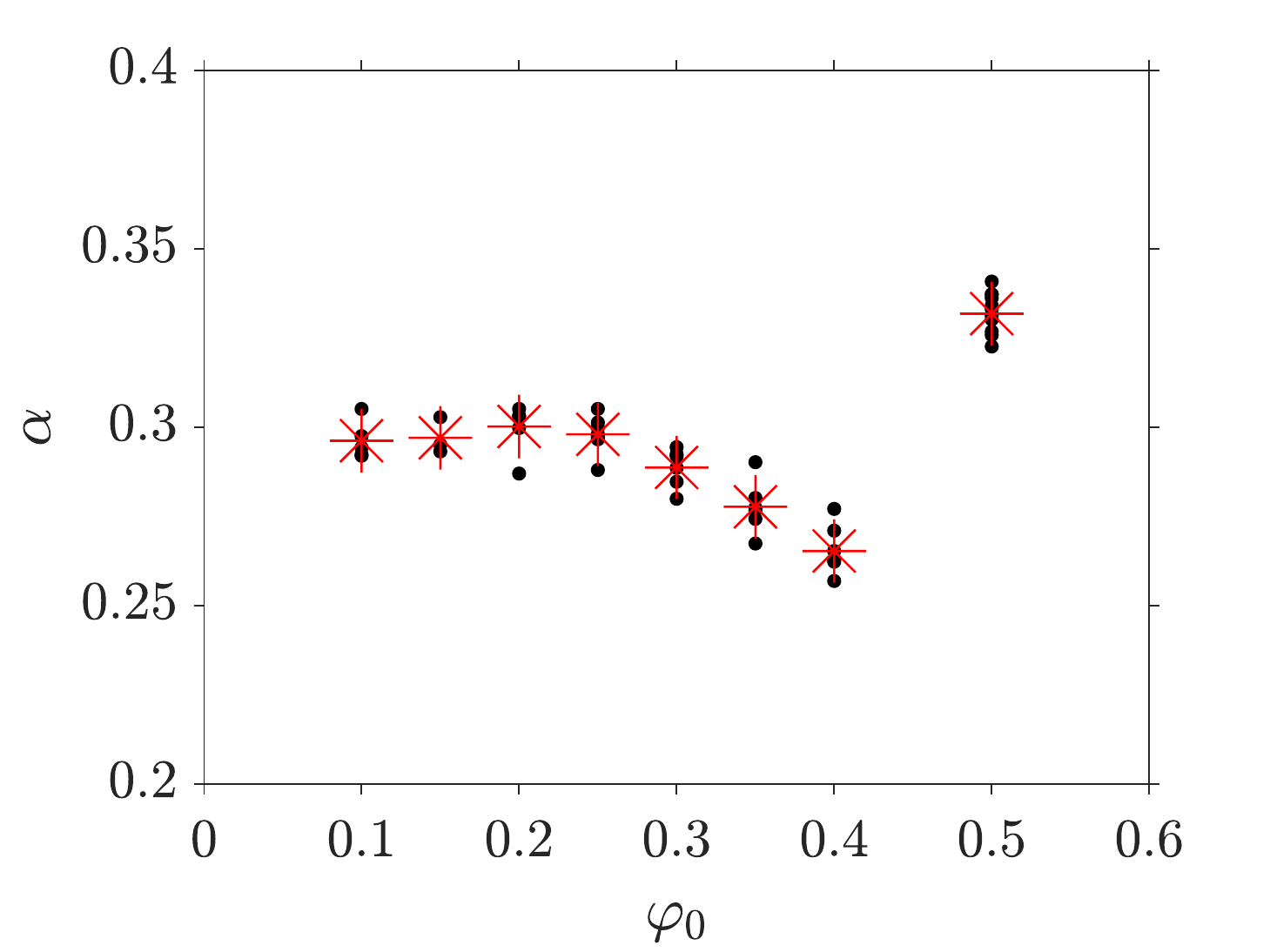}
        \caption{Summary of the obtained coarsening exponents $\alpha$ as a function of the initial mixture composition $\varphi_{0}$ for different quench depths. The red stars indicate the mean values of the obtained exponents.}
        \label{fig:final_result}
    \end{figure}

The morphological evolution for $\varphi_{0}=0.25$, $0.4$, $0.45$ and $0.5$ at constant quench depth $\chi-\chi_{s}=2$ (asterisk markers in Fig.~\ref{fig:testedconfigurations}) is shown in Fig.~\ref{fig:morpho}. The snapshots represent the volume fraction of the second material, from 0 (blue) to 1 (yellow), at different times after phase separation: the snapshots in the first column are taken approximately after one decade of late-stage coarsening ($4\times10^{-5} \ s$), while the snapshots in the second and third column are at two respectively three decades later. 

For the critical mixture, $\varphi_{0}=0.5$, we observe the typical highly interconnected, co-continuous morphology (Fig.~\ref{fig:morpho}(j)-(l)) which persists until the end of the simulation. For both off-critical compositions $\varphi_{0}=0.25$ and $0.4$ (Figs.~\ref{fig:morpho}(a)-(c) and (d)-(f)), the main characteristic is a droplet-in-matrix structure. For $\varphi_{0}=0.25$, all particles or droplets are close to spherical during the whole coarsening process. In contrast, for $\varphi_{0}=0.4$ some of the droplets have a sausage-like shape at the first stage of the coarsening (Fig.~\ref{fig:morpho}(d)). Still, at the end of the simulated coarsening, some droplets deviate from sphericity and have slightly ellipsoidal shapes. The microstructure for the $\varphi_{0}=0.45$-mixture in the first coarsening stage (Fig.~\ref{fig:morpho}(g)) contains mainly co-continuous or worm-like domains. During further coarsening, the co-continuous and worm-like structure elements gradually transform into an ellipsoidal-shaped dominated morphology (Fig.~\ref{fig:morpho}(i)). Note that even after three decades of late-stage coarsening (last column of Fig.~\ref{fig:morpho}), the simulation boxes are still much larger than the domain size, ensuring good statistics for the evaluation of the coarsening kinetics.

Figure~\ref{fig:kstudy_lengthscale} shows the evolution of the characteristic length scale $L(t)$ for the four morphologies presented in Fig.~\ref{fig:morpho} during the late stage of the coarsening. For $\varphi_{0}=0.25$, $0.4$ and $0.5$, the $L$ increases according to power laws up to the latest simulated times so that the data can be nicely fitted with Eq.~(\ref{eq:ripening_allgemein}). The obtained value for the coarsening exponent $\alpha$ is slightly sensitive to the time domain selected for the fit and can slightly differ between two different simulations performed with exactly the same parameters. In order to evaluate the precision on the measurement of $\alpha$, we not only vary the time domain used for the fit, but also perform 10 successive simulations of the same system ($\varphi_{0}=0.3$, $\chi=4$). Altogether, we estimate the standard deviation to be roughly $s=0.0075$.

As it can be seen from Fig.~\ref{fig:kstudy_lengthscale}, the critical mixture coarsens with a coarsening exponent $\alpha=1/3$. This result is in line with the literature as stated in Sec.~\ref{sec:introduction} in Tab.~\ref{tab:literature}. We note that the typical co-continuous morphology is far away from the LSW-assumption but we confirm the LSW-exponent of $\alpha=1/3$ for the critical composition. For the very off-critical system $\varphi_{0}=0.25$, we find the coarsening exponent to be $\alpha=0.3$. For the slightly off-critical composition $\varphi_{0}=0.4$, the coarsening exponent is as small as $\alpha=0.26$, significantly below the initially expected value of $1/3$. This evidence for late-stage coarsening occurring slower than $t^{1/3}$ for off-critical mixture compositions has also been obtained by Garcke et al.\cite{garcke_2003}, by Rogers and Desai\cite{rogers_desai_1989_offcritical} and by Brown and Chakrabarti\cite{chakra_brown_1993}, although newer studies report $\alpha=1/3$ for the off-critical compositions $\varphi_{0}=0.25$ and 0.3 \cite{sheng_coarsening_2010, dai_computational_2016}. 

The comparison of the morphologies (Fig.~\ref{fig:morpho}) and of the coarsening kinetics (Fig.~\ref{fig:kstudy_lengthscale}) suggests an influence of the morphology on the coarsening exponent. The highest value of $\alpha$ is found for the co-continuous morphology ($\varphi_{0}=0.5$, $\alpha=1/3$) while the coarsening is slightly slower for the spherical droplet structure ($\varphi_{0}=0.25$, $\alpha=0.3$). The ellipsoidal-shaped droplet structure produces the slowest coarsening kinetics ($\varphi_{0}=0.4$, $\alpha$=0.26). This hypothesis is supported by the more complex coarsening kinetics of the $\varphi_{0}=0.45$ blend: for $\varphi_{0}=0.45$, a good fit to Eq.~(\ref{eq:ripening_allgemein}) is not possible. We find that the evolution of $L(t)$ is characterized by a transition between two different coarsening exponents. At the beginning of the late-stage coarsening (until approximately $1\times10^{-4} \ s$) the morphology is almost co-continuous or showing worm-like droplets (Fig.~\ref{fig:morpho}(g)), and the structure coarsens with $\alpha=0.314$, which is very close to the value of $1/3$ obtained for the fully co-continuous morphology ($\varphi_{0}=0.5$, Figs.~\ref{fig:morpho}(j)-(l)). At the end of the coarsening (from approximately $1\times10^{-4} \ s$ to the end of the simulation), the morphology is more and more dominated by ellipsoidal-shaped domains (Fig.~\ref{fig:morpho}(i)) and the structure finally coarsens with $\alpha=0.264$. This is similar to the $\varphi_{0}=0.4$ blend (Fig.~\ref{fig:morpho}(f)) for which we obtain $\alpha=0.264$.

The complete overview of the coarsening exponents $\alpha$ obtained for the quenching conditions marked in Fig.~\ref{fig:testedconfigurations} is given in Fig.~\ref{fig:final_result}. In all simulations with $\varphi_{0}$ varying from $0.1$ to $0.4$, and for $\varphi_{0}=0.5$, the morphology coarsens following a power law. On the one hand, at fixed composition, we are not able to observe a significant dependence (within two times the standard deviation) of the coarsening exponent on the quench depth. The red crosses in Fig.~\ref{fig:final_result} are the mean values of the obtained coarsening exponent $\alpha$ for a specific $\varphi$ for all the tested quench depths. On the other hand, as already discussed, there is a strong dependence of $\alpha$ on $\varphi_{0}$. We find that the exponent reaches its maximum value $\alpha=1/3$ for $\varphi_{0}=0.5$, and its minimum value for $\varphi_{0}=0.4$ whereby $\alpha\approx0.26$. For $\varphi_{0}$ smaller than $0.4$, $\alpha$ increases again, reaching an asymptotic value around $0.3$. The co-continuous morphology ($\varphi_{0}=0.5$) seems to evolve faster than the spherical droplet structure ($\varphi_{0}=0.1$ to $0.25$), while the ellipsoidal-shaped droplet structure ($\varphi_{0}=0.3$ to $0.4$) produces the smallest values of $\alpha$. For $\varphi_{0}=0.45$ and for $\varphi_{0}=0.475$, a fit to Eq.~(\ref{eq:ripening_allgemein}) with a single exponent is not possible and we hence exclude these quenching conditions from Fig.~\ref{fig:final_result}. For these compositions, the morphology transitions from almost co-continuous to ellipsoidal droplets, associated with a significant decrease of the coarsening exponent $\alpha$.

In addition, we also evaluate $\alpha$ based on the free energy decrease of the system: similar to the domain sizes, the energy is expected to decrease following a power law. We therefore evaluate and fit the decrease of the interfacial energy for all simulations reported in Fig.~\ref{fig:final_result} and find again composition-dependent coarsening exponents. Interestingly, the $\alpha$-values obtained with this method are systematically about $0.02$ higher than the $\alpha$-values obtained from the evaluation using the structure factor. For $\varphi_{0}=0.1$, we obtain a mean value for $\alpha$ of $0.32$ which is closer to the LSW-prediction of $t^{1/3}$ than the value we found using the characteristic length scale method. For compositions up to $\varphi_{0}=0.25$, corresponding to morphologies made of spherical, isolated droplets, the time evolution of the energy therefore nicely matches the LSW-prediction.

In order to fully understand the coarsening kinetics given by Eq.~(\ref{eq:ripening_allgemein}), we finally need to focus on how the constant $C$ depends on the model parameters. In particular, since the coarsening exponent $\alpha$ does not always have the same value, we also expect $C$ to be $\alpha$-dependent. In the following, we propose an equation for $C$ and check its validity with the help of three arguments.

The first argument is a scaling argument. In our phase-field framework, the equations are written and solved in a dimensionless form using $\hat{G}^\mathrm{loc} = G^\mathrm{loc}/\mu_{0}$, $\hat{G}^\mathrm{nonloc} = G^\mathrm{nonloc}/(\mu_{0}l_{sc}^2)$ for the energies, $\hat{l} = l/l_{sc}$ for the lengths, $\hat{t} = t/t_{sc}$ for the times and $\hat{\Lambda} = \Lambda/D_{sc}$ for the mobilities, with the scaling factors defined as $\mu_{0}=RT/v_0$, $l_{sc}=\sqrt{\kappa/(2\mu_{0})}$, and $t_{sc}=\kappa/(2\mu_{0}D_{sc})$. $D_{sc}$ has the unit of a diffusion coefficient and, since we consider a simple model with constant mobility, can be chosen for instance as $D_{sc}=\Lambda$. Naturally, the equation for the coarsening kinetics can be written in the adimensionalized form, similar to Eq.~(\ref{eq:ripening_allgemein}):
    \begin{equation}
    \label{eq:ripening_adim}
         \hat{L}(t)^{1/\alpha} - \hat{L}(t_{0})^{1/\alpha} = \hat{C} (\hat{t}-\hat{t}_{0}),
    \end{equation}
where $\hat{L}=L/l_{sc}$ is the dimensionless characteristic length and $\hat{t} = t/t_{sc}$, $\hat{t_0} = t_0/t_{sc}$ are the  dimensionless characteristic times. Using the definition of $t_{sc}$ and $l_{sc}$, this leads to
    \begin{equation}
    \label{eq:ripening_scaled}
         L(t)^{1/\alpha} - L(t_{0})^{1/\alpha} = \hat{C} D_{sc} \left(\frac{\kappa}{2\mu_0}\right)^{\frac{1}{2\alpha}-1}(t-t_{0}).
    \end{equation}
As a second argument, we propose that the adimensional constant $\hat{C}$ may vary as $\hat{C} = c^{\frac{1}{\alpha}}\frac{\hat{\Lambda}\hat{\sigma}}{\Delta\varphi_\mathrm{bino}}$, where $c$ is a fitting prefactor, $\hat{\Lambda}$ the dimensionless mobility $\hat{\sigma}$ the dimensionless surface tension and $\Delta\varphi_\mathrm{bino}$ the volume fraction difference between the two separated phases: on the one hand, the dependence on $\hat{\Lambda}$, $\hat{\sigma}$ and $\Delta\varphi_\mathrm{bino}$ is simply inspired by the classical models based on the LSW theory. On the other hand, $\hat{C}$ has also to depend on $\alpha$. Following the idea developed in the previous section that the variation of the coarsening exponent $\alpha$ may only be dependent on the morphology of the blend and neither on the mobility, nor on the quench depth and/or surface tension, we assume that the $\alpha$ dependency only applies to the proportionality factor as $c^{\frac{1}{\alpha}}$, which leads to the proposed equation for $\hat{C}$.

The third argument is the calculation of the surface tension of a diffuse interface using the Van der Waals equation $\sigma=\kappa\int (\frac{d\varphi}{dx})^2dx$, which leads to the scaling $\hat{\sigma}=\sigma/\sqrt{\kappa\mu_0}$. Using this in the equation proposed above for $\hat{C}$ and inserting it in Eq.~(\ref{eq:ripening_scaled}), we finally obtain for the scaled coarsening rate constant
    \begin{equation}
    \label{eq:Cconstant}
         C = \frac{c^{\frac{1}{\alpha}}}{\sqrt{2}}
         \frac{\Lambda\sigma}{\mu_0\Delta\varphi_\mathrm{bino}}\left(\frac{\kappa}{2\mu_0}\right)^{\frac{1-3\alpha}{2\alpha}}.
    \end{equation}
Note that with these hypotheses, the unit of the constant is $m^{1/\alpha}s^{-1}$ as expected and that for $\alpha=1/3$ the LSW law $C\sim{\frac{\Lambda\sigma}{\mu_0\Delta\varphi_{\mathrm{bino}}}}$ is recovered.

\begin{figure}[ht]
        \centering
        \includegraphics[scale = 0.58]{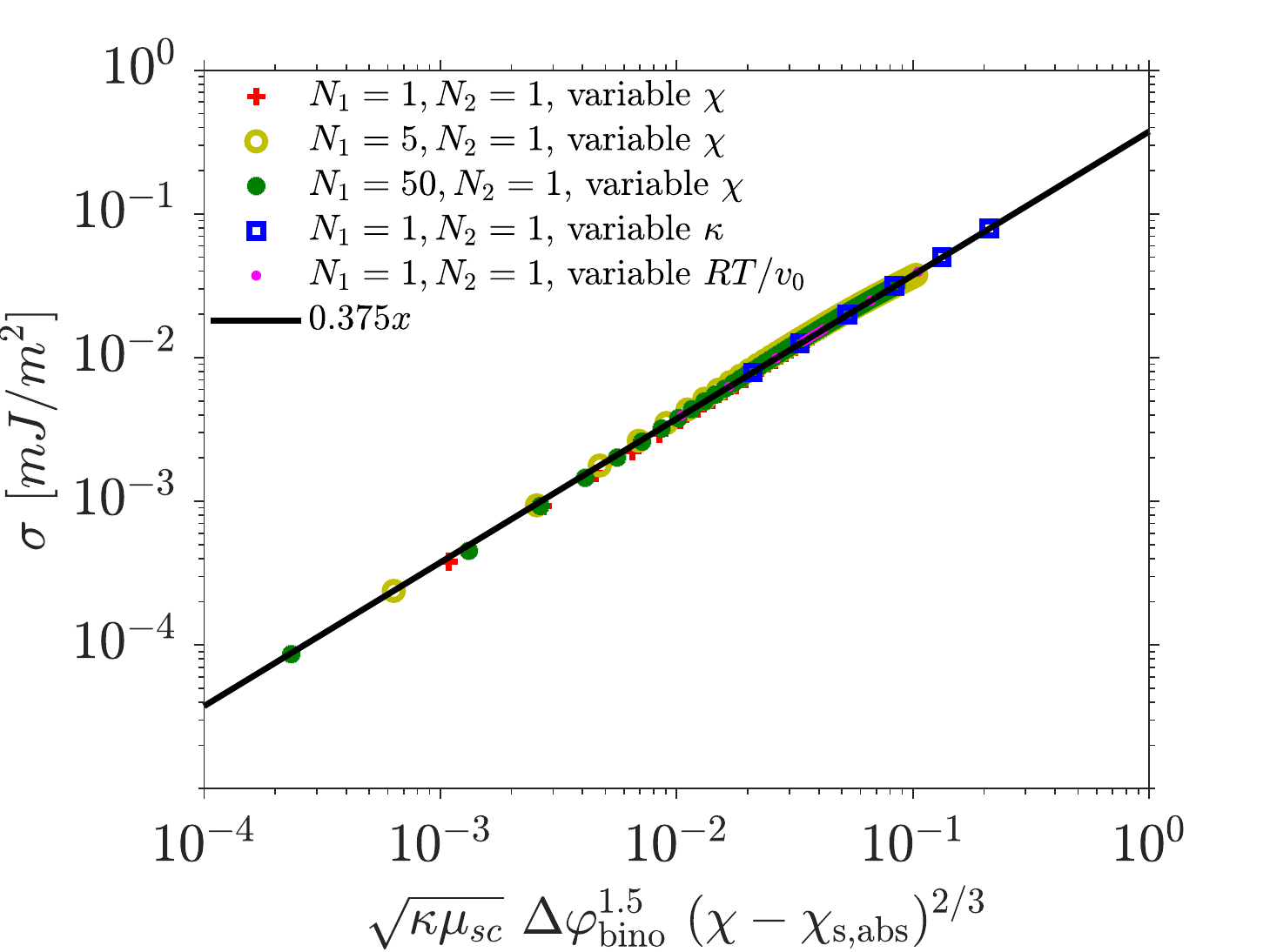}
        \caption{Surface tension obtained from the simulations versus values calculated with Eq.~(\ref{eq:SurfTen}).
        }
        \label{fig:C_result1}
    \end{figure}

In order to compare this equation with the $C$ values obtained from the fit to Eq.~(\ref{eq:ripening_allgemein}), we not only need the binodal composition which can be picked up from the phase diagram, but also the value of the surface tension. To this end, we calculate the one dimensional equilibrium interface profiles between two separated domains, for various parameter sets, varying all the parameters of the Flory-Huggins equation, and calculate the associated surface tension with the Van der Waals formula. We are able to fit the surface tension nicely (see Fig.~\ref{fig:C_result1}) for any binary system with
    \begin{equation}
    \label{eq:SurfTen}
        \sigma=0.375\sqrt{\kappa\mu_0}{\Delta\varphi_\mathrm{bino}}^{1.5}(\chi-\chi_\mathrm{critabs})^{2/3}.
    \end{equation}
    
\begin{figure}[ht]
        \centering
        \includegraphics[scale = 0.58]{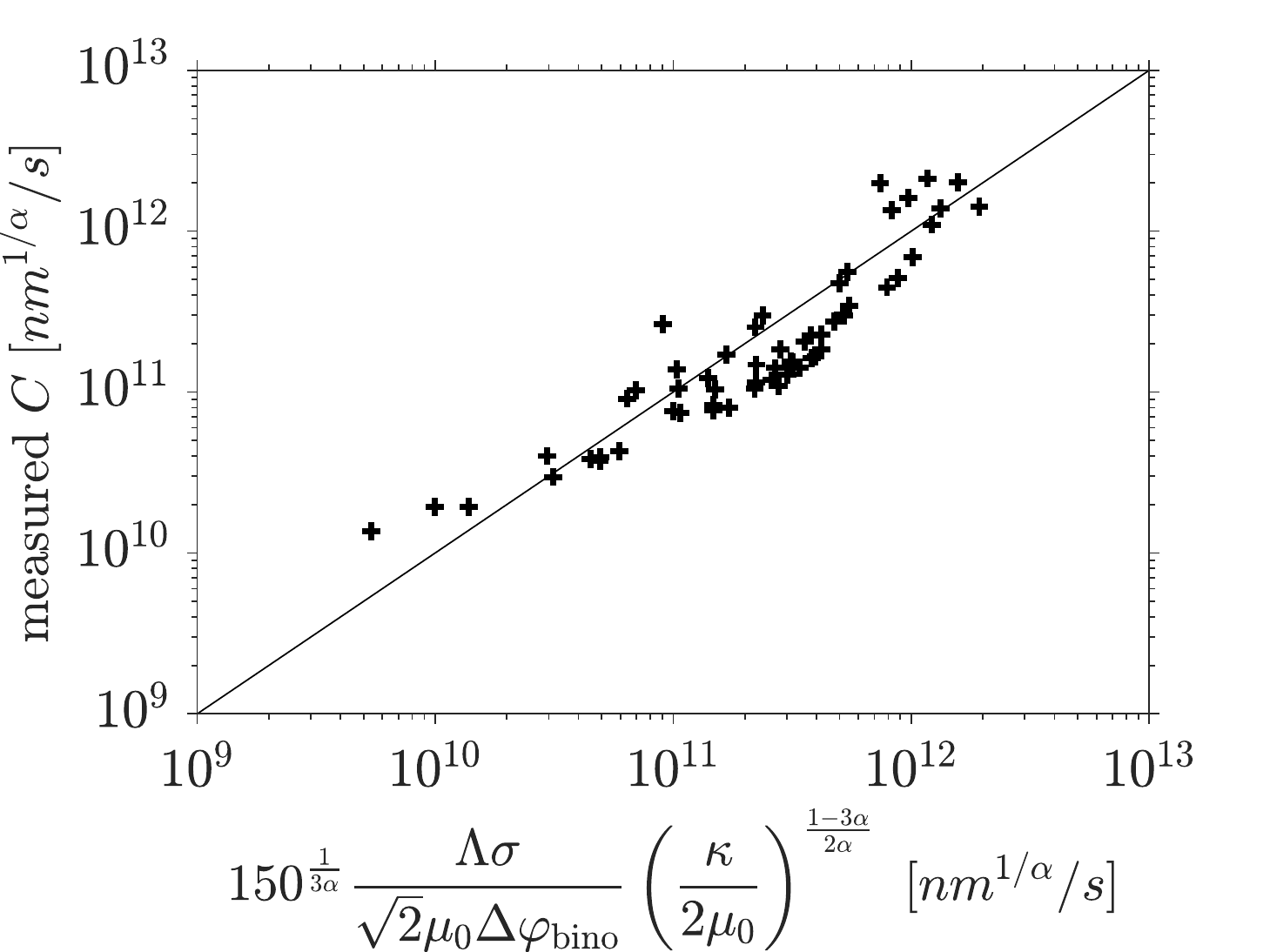}
        \caption{Coarsening rate constants $C$ obtained from the simulations versus values calculated with Eq.~(\ref{eq:Cconstant}).
        }
        \label{fig:C_result2}
    \end{figure}

The comparison between the coarsening constants obtained from the fit of the time-dependent characteristic length scale on the one side, and Eq.~(\ref{eq:Cconstant}) on the other side, is shown in Fig.~\ref{fig:C_result2}. Once again, the data for which the late stage cannot be fitted with a power law (namely $\varphi_{0}=0.45$ and $\varphi_{0}=0.475$) are not taken into account, but apart from that, all simulations reported in Fig.~\ref{fig:final_result} are also used for Fig.~\ref{fig:C_result2}. Despite of some significant deviations, the measured coarsening rate constant can be very nicely predicted with Eq.~(\ref{eq:Cconstant}) with a value of $c=150$, for all initial volume fractions $\varphi_{0}$ and quench depths. 

At the end, all the parameters used in Eq.~(\ref{eq:ripening_allgemein}) ($t_0$, $L_0$, $C$, $\alpha$) are known for systems investigated here (symmetric blends with constant mobilities in two dimensions). This means that the coarsening kinetics can be predicted without the need of any further simulation using Eqns.~(\ref{eq:lsa7}), (\ref{eq:lsa6}), (\ref{eq:SurfTen}), (\ref{eq:Cconstant}) and the $\alpha$ values from Fig.~\ref{fig:final_result}.

\section{Conclusion}
\label{sec:conclusion}

In this paper, we simulated the spinodal decomposition and subsequent coarsening of immiscible binary blends in 2D, using conserved Cahn-Hilliard dynamics together with the Flory-Huggins free energy of mixing. 

First, we simulated the early stages of SD for a broad range of parameters, including asymmetric blends and highly asymmetric, composition-dependent mobility functions. We could check that the time required for phase separation and the initial characteristic size matches the theoretical predictions. 

Second, we investigated the late stage coarsening dynamics of a symmetric blend with a constant mobility. The quench depth and blend composition were varied systematically. As expected, we found that the growth of the characteristic length scale follows a power law. We found no systematic and clear dependence of $\alpha$ on the quench depth. However, the coarsening exponent was found to be composition-dependent, starting from $\alpha=0.30$ for strongly off-critical mixtures, with a minimum of about 0.26 for 40:60 blends and reaching a maximum of $1/3$ for the critical mixture. We propose that the morphology of the phase-separated mixture might be the reason for these variations of the coarsening kinetics, the co-continuous morphology leading to the highest values of $\alpha$, while the smallest values are found for morphologies dominated by ellipsoidal droplets. In addition to the data for $\alpha$, we proposed a model (Eq.~(\ref{eq:Cconstant})) to predict the coarsening rate constant $C$. All these results finally allow the prediction of the late-stage coarsening kinetics using Eq.~(\ref{eq:ripening_allgemein}) for symmetric blends under the assumption of a constant mobility $\Lambda$, without the need of additional simulations.

To shed more light on the observed $\varphi_{0}$-dependence of the coarsening exponent $\alpha$, we plan to quantitatively analyze the obtained morphologies regarding their interfacial shape distribution, mean curvature and other structure metrics according to Refs.\cite{voorhees_2017, andrews_simulation_2020}. This also includes the characterization using Minkowski descriptors as proposed by Manzanarez et al.\cite{MANZANAREZ_2017, MANZANAREZ_2021}. Furthermore, we plan to investigate asymmetric material systems in near future, as well as systems with more sophisticated mobility functions in order to extend, confirm and enrich the results on the coarsening behavior of binary blends.

\begin{acknowledgments}
The authors acknowledge financial support by the German Research Foundation DFG (project HA4382/14-1). The data that support the findings of this study are available from the corresponding author upon reasonable request.
\end{acknowledgments}



%

\end{document}